\documentclass[fleqn,usenatbib]{mnras}

\usepackage{newtxtext,newtxmath}
\usepackage{gensymb}
\usepackage[T1]{fontenc}
\usepackage{ae,aecompl}

\usepackage{graphicx}	% Including figure files
\usepackage{amsmath}	% Advanced maths commands
\usepackage{amssymb}	% Extra maths symbols
\usepackage[utf8]{inputenc}
\newcommand{\Teff}   {$T_{\rm eff}$~}

\newcommand{\logg}{$\log g$} 
\newcommand{\feh}   {$[\mathrm{Fe/H}]$~}

\newcommand{\Prot}   {$P_{\rm rot}$~}
\newcommand{\Psini}   {$P_{\rm rot}/\sin i$~}

\newcommand{\tess} {\textit{TESS}}
\newcommand{\lrlhkt}{\ensuremath{\log R^\prime_{\rm HK} (T_{\rm eff})}}
\newcommand{\lrlhkb}{\ensuremath{\log R^\prime_{\rm HK} ({\rm B-V})}}

\title{The ancient main-sequence solar proxy HIP 102152 unveils the activity and rotational fate of our Sun}

\author[Ancient solar proxy HIP 102152 unveils the fate of our Sun]{Diego Lorenzo-Oliveira,$^{1}$\thanks{E-mail: diegolorenzo@usp.br}
Jorge Meléndez,$^{1}$
Geisa Ponte,$^{2}$
Jhon Yana Galarza$^{1}$
\\
$^{1}$Universidade de S\~ao Paulo, Departamento de Astronomia do IAG/USP, Rua do Mat\~ao 1226, Cidade Universit\'aria, 05508-900 \\S\~ao Paulo, \ SP, Brazil \\
$^{2}$Centro de Radioastronomia e Astrof\'isica Mackenzie, Universidade Presbiteriana Mackenzie,  
Rua da  Consola\c{c}\~ao 930, 01301-000 \\ S\~ao Paulo, SP, Brazil.\\}
\date{Accepted XXX. Received YYY; in original form ZZZ}

\pubyear{2020}

\hypersetup{draft}
\begin{document}
\label{firstpage}
\pagerange{\pageref{firstpage}--\pageref{lastpage}}
\maketitle

\begin{abstract}
We present a detailed analysis of the possible future Sun's rotational evolution scenario based on the 8 Gyr-old solar twin HIP 102152. Using  HARPS high-cadence observations (and \tess\ light curves), we analyzed the modulation of a variety of activity proxies (\ion{Ca}{II}, \ion{H}{I} Balmer, and \ion{Na}{I} lines), finding a strong rotational signal of 35.7 $\pm$ 1.4 days ($\log B_{\rm factor}\sim70$, in the case of \ion{Ca}{II} K line). This value matches with the theoretical expectations regarding the smooth rotational evolution of the Sun towards the end of the main-sequence, validating the use of gyrochronology after solar age.

\end{abstract}


%%%%%%%%%%%%%%%%%%%%%%%%%%%%%%%%%%%%%%%%%%%%%%%%%%

%%%%%%%%%%%%%%%%% BODY OF PAPER %%%%%%%%%%%%%%%%%%

\section{Introduction}

How will the Sun evolve in the future?  In the case of stellar rotation, thanks to seminal works in the past \citep[e.g.][]{skumanich72, barnes03}, we now have a fundamental tool that traces many astrophysical effects \citep[e.g., magnetic fields, interaction with close planets, structural evolution;][]{vidotto14,ferrazmello15} and, as a byproduct of all these correlations with time, stellar rotation can be used as an elegant and straightforward chronometer \citep{barnes07, mamajek08}.

However, stellar rotation as a predictive tool still lacks strong observational constraints in critical regions of its parameter space, for example, old stars with global properties similar to those of the Sun ($\gtrsim$4 Gyr, $\sim$ 1 $M_\odot$, \feh$\sim$ 0.00). Therefore, due to this scarcity, after this age, different theoretical assumptions of radius dependence \citep{reiners12}, angular momentum loss \citep{matt15}, magnetic field topology \citep{see19}, stellar differential rotation profile \citep{benomar18}, for example, might significantly affect the predictions for rotational evolution of the Sun  \citep[][the latter referred hereafter as LO19]{barnes07,donascimento13,donascimento14,lorenzo19}. As suggested by \cite{vansaders16} (hereafter VS16), around solar age or a critical value of Rossby number $\sim$ 2 \citep[Ro $\equiv$ \Prot/$\tau_{\rm conv}$, where $\tau_{\rm conv}$ is the convective turnover time;][]{noyes84} the efficiency of magnetic braking may be significantly reduced, leading to a dramatically different rotational evolution scenario for the Sun. Therefore, it is crucial to constrain this age interval with new observations. To solve this burning conflict, and assess how solar-mass \& solar-metallicity stars cease to slow down after solar age, we provide unique constraints to unveil the rotational fate of our Sun towards the end of the main-sequence through an extensive activity analysis of the ancient solar twin HIP 102152.
\begin{table}
\caption{Main stellar parameters derived for HIP 102152. [1] \citet{gaia18}; [2] \citet{spina17}; [3] \citet{lorenzo18}; [4] LO19; [5]
\citet{lorenzo16};
[6] This work; [7] \citet{barnes10}; [8] \citet{mittag13}; [9] \citet{carlos19}}
\begin{center}
\begin{tabular}{c c c}
\hline
Parameter & Value & Reference \\
\hline
\hline
\textit{Alias} &  & \\ 
Gaia DR2 & 6799090223510802176 & [1] \\
HD & 197027 & $--$ \\
TIC & 212354618 & $--$\\  
\hline
\textit{Atmospheric parameters} &  & \\ 
\Teff [K] & $5718 \pm 4$ & [2] \\
\feh & $-0.016 \pm 0.003$ & [2]\\
\logg & $4.325 \pm 0.011$ & [2]\\
\hline
\textit{HR diagram} &  & \\ 
mass [M$\odot$] & $0.97 \pm 0.03$ & [4] \\
radius [R$\odot$] & $1.07 \pm 0.02$ & [4]\\
\hline
\textit{Stellar Activity} &  & \\
$<S_{\rm MW}>$ & $0.159 \pm 0.001$ & [6] \\
\lrlhkb & $-5.01 \pm 0.02$  & [6] \\
\lrlhkt & $-5.12 \pm 0.02$ & [6] \\
$\log R^+_{\rm HK}$ & $-5.2$ & [6][8] \\
\hline
\textit{Stellar Rotation} &  & \\
\Psini [d] & $34.8 \pm 4.7$ & [4] \\
\Prot (\tess) [d] & $>27$  & [6] \\
$P_{\rm rot , TESS, P/\sin i}$ [d] & $34.4_{-6.2}^{+ 6.9}$  & [6] \\
\Prot  [d] (adopted) [d] & $35.7 \pm 1.4$ & [6] \\
\hline
\textit{Stellar Age} &  & \\
$t_{\rm iso}$ [Gyr] & $8.0_{-0.4}^{+0.3}$ & [4]\\
$t_{\rm HK~1}$ [Gyr] & $7.6 \pm 1.0$ & [3][6] \\
$t_{\rm HK~2}$ [Gyr] & $6.6^{+2.5}_{-1.8}$ & [5][6] \\
$t_{\rm Gyro}$ [Gyr] & $7.6\pm0.6$ & [6][7] \\
$t_{\rm Li}$ [Gyr] & $9.1 \pm 1.1$ & [6][9] \\
$<t_{\rm Chem}>$ [Gyr] & $7.2 \pm 1.0$ & [2][6] \\ 
\hline
\hline

\end{tabular}
\end{center}

\label{table:id}
\end{table}
\section{Observations, Parameters and Activity Indicators}

The star HIP 102152, located at $\sim$ 78.4 parsecs from us, is one of the most interesting solar twins ever discovered \citep[][hereafter M13]{monroe13}. Its macroscopic characteristics are nearly indistinguishable from what we would expect for the ancient Sun. Considering its isochronal age, mass and metallicity ($t_{\rm iso}$ = $8.0_{-0.4}^{+0.3}$ Gyr, 0.97$\pm$0.03 M$_\odot$, \feh = $-$0.016$\pm$0.003 dex, LO19), HIP102152 is placed at almost the end of the main-sequence. From a chemical point of view, it has the same level of deficiency of refractory elements as the Sun, which could be a hint about the formation of eventual terrestrial planets (M13). Additionally, empiric age indicators such as chromospheric activity \citep{lorenzo16a, lorenzo18}, Li abundances \citep{carlos19}, [Y/Mg], and [Y/Al] \citep{spina17} also yield ages ranging from 7 to 9 Gyr, which agrees with the estimates from \citet{lorenzo19}. In Table \ref{table:id} we summarize the main stellar parameters derived for HIP 102152.

The observations were carried out for \textit{exactly} 10 years, between 05-22-2009 and 05-22-2019 (MJD=54973.9$-$58625.8), with the HARPS spectrograph \citep{mayor03} fed by the 3.6 m telescope at La Silla Observatory, under the programs 183.D-0729 (PI: Bazot, M.), 292.C-5004 (PI: Mel\'endez, J.), 188.C-0265 (PI: Mel\'endez, J.), 0100.D-0444 (Lorenzo-Oliveira, D.), and 0103.D-0445 (Lorenzo-Oliveira, D.). 

To perform the activity analysis, we selected only observations with a signal-to-noise ratio greater than 30 around the Ca II lines and constrained our sample to those observations with angular separation greater than 15 degrees away from the Moon to avoid excessive contamination of scattered light. Our sample is composed of 52 spectra, of which 34 were taken at high cadence during a short observational window over $\sim$4 months (MJD=56484.69$-$56625.50). Different activity indicators were used in this work such as \ion{Ca}{II} H (3933.663$\pm$0.595 \AA) and K (3968.469$\pm$0.595 \AA) lines; Balmer lines H$\alpha$ (6562.8$\pm$0.5 \AA), H$\beta$ (4861.32$\pm$0.5 \AA), H$\gamma$ (4340.46$\pm$0.4\AA), H$\delta$ (4101.76$\pm$0.25 \AA), and H$\epsilon$ (3970.07$\pm$0.3\AA); \ion{Na}{I} lines (5889.95$\pm$0.25 \AA, 5895.92$\pm$0.25 \AA). We defined their respective pseudo-continuum regions based on \cite{maldonado19} and \cite{giribaldi19}, albeit with slight modifications: 3901.07 and 4001.07 \AA\ ($\Delta\lambda$ = 20 \AA; \ion{Ca}{II} and H$\epsilon$), 6500.625 and 6625.55 \AA\ ($\Delta\lambda$ = 5 \AA; H$\alpha$), 4845.0 and 4880.0 \AA\ ($\Delta\lambda$ = 10 \AA; H$\beta$), 4318 and 4366 \AA\ ($\Delta\lambda$ = 10 \AA; H$\gamma$), 4085 and 4120 \AA\ ($\Delta\lambda$ = 20 \AA; H$\delta$), 5845.0 and 5940 \AA\ ($\Delta\lambda$ = 5 \AA; \ion{Na}{I}). Instrumental activity indices are given by the ratio between the sum of the line fluxes and their respective continuum regions. Internal errors are estimated propagating the typical photonic errors in each HARPS spectral order analyzed. In the case of \ion{Ca}{II} lines, we also use the S index converted into the Mount Wilson scale \citep[see][]{lorenzo18}. To calculate the activity indices, we resampled the spectra using linear interpolation, assuming constant steps of 0.01 \AA. Then, for a given spectral region (defined by each chromospheric indicator), we build a master spectrum based on a combination of all available observations and, subsequently, normalize each one of the observations in comparison to its master spectrum, to correct minor continuum variations from one observation to another. Spectra taken within a 1-day interval were combined to improve the S/N.

\section{Rotational period of HIP102152}

We started the calculation of the most likely rotational period of HIP 102152 using the \ion{Ca}{II} S index, which is well-known to be strongly correlated to stellar rotation \citep{mamajek08}. The star is very inactive and shows a low level of chromospheric variability over 10 years of monitoring, possibly indicating a rotation level around 30$-$40 days, according to the \citet{mamajek08} rotation$-$activity calibration. As a starting point, we analyzed the activity time series of HIP 102152 during the highest cadence dates (MJD = 56484.70$-$56625.51). We use \textit{Generalized Lomb-Scargle} periodogram (GLS) analysis to detect a clear peak around $\sim$36 days. From this initial guess, we refined our analysis by using the Gaussian Process regression fit to deal with quasi-periodic (QP) trends in the activity time series \citep{haywood15}. For a given chromospheric indicator \textit{i}, we define an appropriate combination of covariance functions relating to different epochs ($t$ and $t'$) of observations to build our QP activity model:
\begin{equation}
    k(t,t')\equiv \mathcal{I}_{\rm const}+\mathcal{A}\exp\left(-\frac{||t-t'||}{2 \ell^2} -\Gamma\sin^2\left[\frac{\pi}{ P_{\rm rot}}||t-t'||\right] \right)+\sigma^2\delta_{t,t'},
\end{equation}
where $\mathcal{I}_{\rm const}$ gives a constant scale to match the observed mean activity level of the star, $\mathcal{A}$ is the amplitude of rotation signal, $\ell$ is commonly interpreted as the timescale of rising and decay of active regions. The harmonic nature of the time series is represented by $\Gamma$, and the white noise term is $\sigma^2\delta_{t,t'}$. We adopted log-normal prior distribution for the hyperparameters: $\mathcal{I}_{\rm const}$ ($\mu = {\rm <i>}$,$\sigma = \sigma_{\rm <i>}$), \Prot ($\mu = P_{\rm rot, GLS}$,$\sigma = 0.2\times P_{\rm rot, GLS}$), and $\ln\Gamma$ \citep[$\mu = -2.3$, $\sigma = 1.4$, as in][]{angus18}. For the other hyperparameters, we adopted Jeffrey's prior. To find the optimal solution and the associated errors for the stellar rotation GP model, we use the \textit{emcee} \citep{foremanmackey13} \textit{Python} implementation of the affine-invariant ensemble sampler for \textit{Markov Chain Monte Carlo} method (MCMC) following \citet{angus18}. In brief, we start the MCMC process with 100 walkers spread around an optimal solution obtained by maximum likelihood optimization. Then, we evaluate the convergence of chains every 100 steps, checking the chain autocorrelation timescale ($\tau_{\rm chain}$) and the consistency of walkers solutions through Gelman$-$Rubin statistics (\^R).  We define as a convergence criterion when  $\tau_{\rm chain}$ is less than 10\% of the total chain length, $\tau_{\rm chain}$ is stable concerning the previous chain evaluation within 1\%, and \^R less than 1.03. We then discard the initial iterations (3$\times$ $\tau_{\rm chain}$) and randomly resample 5000 samples to represent our final estimate of posteriori probability distribution. 

\begin{figure}
\centering
  \begin{minipage}[t]{0.49 \linewidth}
\centering
    \resizebox{\hsize}{!}{\includegraphics{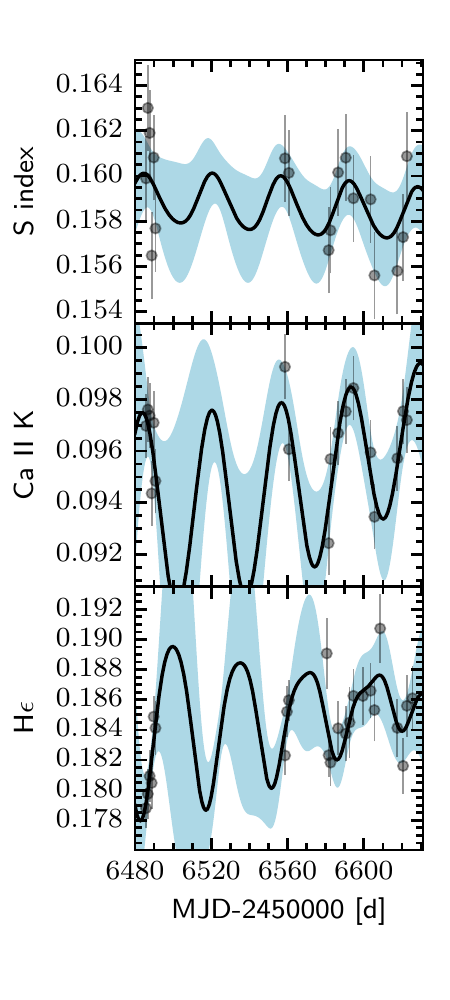}}
  \end{minipage}
  \begin{minipage}[t]{0.49 \linewidth}
\centering
    \resizebox{\hsize}{!}{\includegraphics{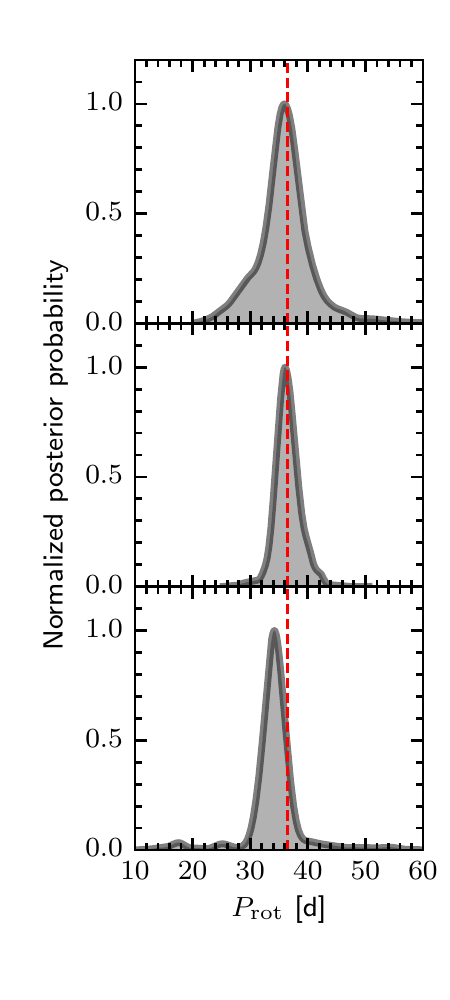}}
  \end{minipage}
  \caption{\textit{Left panel:} Rotational signal detected in S index, \ion{Ca}{II} K, and H$\epsilon$ lines (upper, middle, and lower panels, respectively). Solid lines and shaded blue regions are QP model and 2$\sigma$ predictions. The error bars are given by the quadratic propagation of white noise and spectroscopic index errors. \textit{Right panel:} \Prot pdf for the same activity indicators. Red dashed lines is the gyrochronology prediction from \citet{barnes10}.}
\label{fig:S}
\end{figure}

We found for the \ion{Ca}{II} S index a rotational period of 36.1$_{- 3.3}^{+ 3.8}$ days. Alternatively, we built another GP model composed of only the constant kernel $\mathcal{I}_{\rm const}$ that relates the mean activity level in order to assess its statistical relevance over the QP model. In Fig. \ref{fig:S} (upper left panel), we show the rotational behavior found for the S index. The blue shaded area is our $\pm2\sigma$ QP model prediction. The probability distribution of \Prot is shown in the right panel. The red dashed line is the gyrochronology prediction for 8 Gyr-old solar mass star using \citet{barnes10} relations. The ratio between MCMC QP and Constant model posterior probabilities ($\log p_{\rm QP}/p_{\rm Const}\equiv\log B_{\rm factor}$) gives us an idea about which model is more suitable to describe the data. The greater the $B_{\rm factor}$, the greater the probability that we should favor the QP model. According to \citet{kass95}, $\log B_{\rm factor}\geq2$ has a \textit{decisive} probability favoring QP model. For the S index, we found a value of $\log B_{\rm factor}\sim60$. The same result is obtained by analyzing each \ion{Ca}{II} line, now adopting as a \Prot\ prior the S index $p_{\rm QP}$ distribution. As seen in Fig. \ref{fig:S} (panel immediately below the S index one), the modulation of the \ion{Ca}{II} K line is even more significant than those obtained for the S index, reaching $\log~B_{\rm factor}\geq70$ for \Prot = $36.1_{-2.0}^{+2.6}$ days. Although the lower $B_{\rm factor}$ compared with \ion{Ca}{II} K line and S index, \ion{Ca}{II} H shows consistent rotational period. For the other indicators, we repeat the same strategy used for the S index. In the lower panel of Fig. \ref{fig:S}, we also show the H$\epsilon$ line performance. %as rotation discriminator.
In Table \ref{table:prots}, we show the best indicators: \ion{Ca}{II} K line, \ion{Ca}{II} S index (H+K), \ion{Ca}{II} H line, and H$\epsilon$. The remaining set of indicators that show $\log B_{\rm factor} < 2$ are followed by H$\gamma$, H$\alpha$, \ion{Na}{I} D2 and D1 lines, H$\delta$, and H$\beta$. As we decrease $B_{\rm factor}$, the results begin to spread out over a wider range of \Prot possibilities (with some of them yielding multimodal \Prot estimates), but still within a given range (\Prot = $31.1_{-3.2}^{+4.6}$ days) that is statistically indistinguishable from the values provided by the best indicators (\Prot = $35.7$ $\pm$ $1.4$ days). Therefore, we interpret that, as we consider more reliable estimates of \Prot, the rotational signal of HIP 102152 becomes progressively stronger, converging into a sharp distribution of possibilities that peak at $\sim$36 days. 

\begin{figure}

  \begin{minipage}[t]{1 \linewidth}
\centering
    \resizebox{\hsize}{!}{\includegraphics{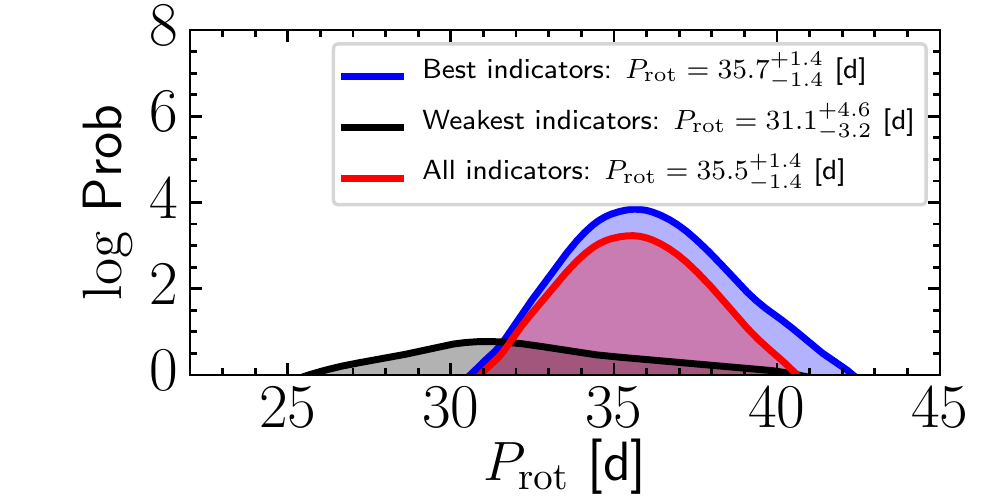}}
  \end{minipage}
  \begin{minipage}[t]{1 \linewidth}
\centering
    \resizebox{\hsize}{!}{\includegraphics{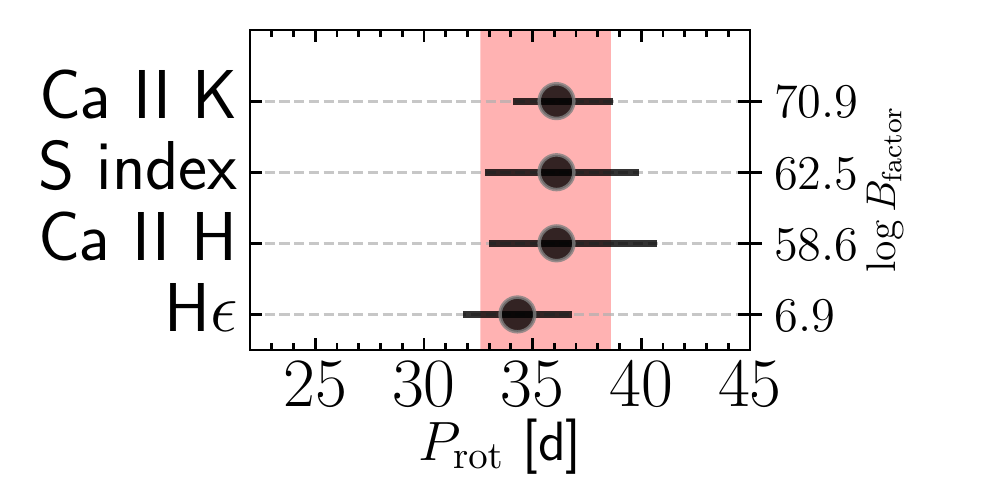}}
  \end{minipage}
  \caption{\textit{Left panel:} joint \Prot probability distribution for 3 combinations of indicators: best (in blue), all (red), and weakest (black) indicators. \textit{Right panel:} \Prot mode and errors for each one of the best indicator as a function of its QP model statistical significance in comparison to Constant model. The red shaded region is the 2$\sigma$ region around 35.7 days.}
\label{fig:rot_indicators}
\end{figure}

In Fig. \ref{fig:rot_indicators}, left panel, we show the $p_{\rm QP}$ of chromospheric indicators. Three scenarios were analyzed: the first one (in blue), we combined the $p_{\rm QP}$ of the best indicators ($\log B_{\rm factor} > 2$). The second one (in red) is given by the combination of all indicators used in this work. The last scenario (in black) is  based on the weakest rotation indicators derived for HIP 102152 ($\log B_{\rm factor} \leq 2$). In Fig. \ref{fig:rot_indicators} (right panel), we summarize the results found for each indicator. The mode of each posterior distribution gives the centroid of the error bars, and the width of the same distribution represents the error bars at 60\% of its height. The red shaded region shows the 95\% confidence interval for \Prot estimates using the best indicators (\Prot = 35.7 $\pm$ 2.8 days, $\pm2\sigma$). 
\begin{table}
\caption{Rotation periods derived for \ion{Ca}{II} lines and H$\epsilon$ according to the QP model. $B_{\rm factor}$ is given by the ratio between QP and Constant model posteriors.}
\begin{center}
\begin{tabular}{c c c}
\hline
 Indicator & \Prot & $\log B_{\rm factor}$\\
\hline
 \ion{Ca}{II} K & 36.1$_{- 3.3}^{+ 3.8}$  & $+$70.91 \\
 \ion{Ca}{II} S (H+K) & 36.1$_{- 2.0}^{+ 2.6}$  & $+62.54$ \\
 \ion{Ca}{II} H & 36.1$_{- 3.1}^{+ 4.6}$  & $+58.63$ \\
H$\epsilon$ & 34.3$_{- 2.5}^{+ 2.5}$  & $+6.94$\\
\hline
\end{tabular}
\end{center}
\label{table:prots}
\end{table}
The star HIP 102152 was monitored photometrically in short cadence mode (2 minutes) by \tess\ mission along with its sector 1 (camera 1, CCD 4, between MJD = 58325.3 and 58353.2 days). We extracted the light curve from target pixel files using the \textit{lightkurve} python package \citep{lightkurve18} and corrected the long-term pointing jitter using \textit{Pixel-Level Decorrelation} method \citep[PLD,][]{luger16}. Because of the restricted time span of 27 days, we could only assign an upper limit for HIP 102152 variability, which is $>$ 27 days. Still, it is possible to go one step further, bracketing the photometric information from one side and the \Psini from the other (see Table \ref{table:id}). To do so, the corrected light curve was binned in chunks of 10 h and then, we applied our QP model, now with an additional \Prot prior given by the Survival function $\mathcal{S}$($\mu \equiv$\Psini$=34.8$ and $\sigma = 4.7$ days) resulting in $P_{\rm rot , TESS, P/\sin i} =  34.4_{-6.2}^{+ 6.9}
$ days ($\log B_{\rm factor} = 21$). 

\section{Future of the Sun: HIP 102152 in the context of magneto-rotational evolution}

With $\sim$36 days of rotational period and $\sim$8 Gyr, HIP 102152 lies at a critical region, where the evolutionary speed increases and the physical effects accumulated over the entire stellar lifetime become more relevant, such as radius dependence \citep{reiners12}, mass loss rate \citep{fionnagain18}, magnetic field geometry \citep{petit08}, and metallicity \citep{lorenzo16,amard20}. Therefore, it is possible to constrain rotational models better and start to unveil the solar evolutionary path within $\sim$3.5 Gyr from now. 
 
After the solar age, solar mass stars become very inactive and with their surfaces dominated by plages \citep[unlike the younger counterparts dominated by spots,][]{hall09}. The balance between plages and spots can introduce different signatures in the light curves as we vary the stellar metallicity \citep{witzke20}. Besides the possible systematic trends in light curve extraction techniques \citep{cui19}, this spot/plage balance hampers the \textit{true} \Prot detection \citep{reinhold19} and also makes the interpretation of the light curves not straightforward, requiring sophisticated approaches \citep[e.g.][]{amazogomez20}. On the other hand, spectroscopy as a benchmark technique is a good alternative because it carries a wealth of important physical information \citep{strassmeier18}, as a large number of activity tracers throughout the optical to near-infrared spectral regions. In the case of \ion{Ca}{II} lines, for example, they are collisionally controlled set of lines formed in the lower chromosphere. Besides, its response to the increase of plage coverage \citep{meunier09} makes it an interesting rotational discriminator for \textit{plage-dominated} old stars.

It is expected that close-in massive planets \citep[or engulfed planets,][]{melendez17} can significantly alter the stellar rotational evolution \citep{ferrazmello15} for a considerable amount of time. We investigated these possibilities from planetary engulfment point of view and possible presence close-in massive planets. This star has the same abundance pattern as a function of the condensation temperature in comparison to the Sun (M13), showing no sign of enhanced refractories due to planet engulfment. Nearby massive companions are also unlikely since the radial velocity variations over the time span of observations are less than a few m/s, which excludes the possibility of both binarity and the presence of nearby massive planets. 

Therefore, we can consider HIP 102152 as a genuine old \textit{solar proxy} that may give us an exciting chance to understand the future of the Sun in many ways. From the magnetic evolution side, during the 10 years of observations, its low level of activity never reached the solar activity typical level. This corroborates the scenario of activity evolution towards the end of the main-sequence \citep[see][]{lorenzo18}. Possibly, HIP 102152 approached to the basal level of activity. According to \citep[][hereafter M18]{mittag18}, when the star's activity converges into the basal flux level, it is reasonable to expect a change in the dominant components of the stellar dynamo (from a vigorous global magnetic field to small-scale turbulent components). M18 points out that this stage would occur around Ro $\sim$ 1. In other words, the dynamo of stars with slow rotation will no longer be dominated by rotational effects. We use the activity and $\tau_{\rm conv}$ relations found in M18 to be 39 $\pm$ 6.3 days (value close to \Prot = 35.7) and Ro = 0.91 $\pm$ 0.15, respectively for HIP 102152. The chromospheric activity corrected from basal component is $\log R_{\rm HK}^{+}$ $\sim$ $-$5.2 \citep[$\log R_{\rm HK, \odot}^{+}\sim-5.06$,][]{mittag13}. For comparison, this star lies in the lower part of the activity$-$Ro diagram shown in M18. Based on these values, we estimate that HIP 102152 will reach Ro = 1 in less than 0.5 Gyr from now (according to the models used in LO19), a negligible age interval that makes it statistically within the expected region of basal flux dominance. In brief, LO19 built \Prot tracks from modified Kawaler wind-law and  \textrm{YaPSI} models \citep{spada17}. LO19 assumed the dominance of structural effects over magnetic braking terms whenever the star approaches into the turn-off region or a given Rossby number ($Ro_{\rm crit}$), leaving only the moment of inertia to drive the subsequent \Prot evolution.

\begin{figure}
\centering
  \begin{minipage}[t]{1 \linewidth}
\centering
    \resizebox{\hsize}{!}{\includegraphics{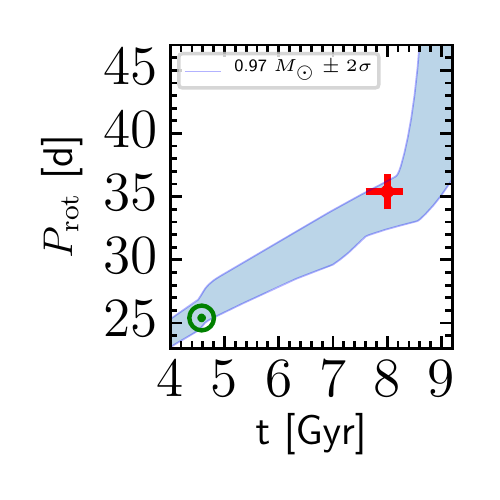}}
  \end{minipage}
  \caption{Age-Rotation diagram with model predictions for 0.97 M$_\odot$ star using modified Kawaler wind-law and  \textrm{YaPSI} tracks \citep{spada17}. The red cross is HIP 102152, and the Sun is represented by its usual symbol. The shaded region is the 2$\sigma$ prediction band.}
\label{fig:rot_age}
\end{figure}

Alternatively, around $\sim$4 Gyr or a critical value of Ro (Ro$_{\rm crit} \sim2$,  note that the $\tau_{\rm conv}$ here is in a different scale to those obtained by M18), it is hypothesized that stellar rotation will no longer be sufficient to maintain an organized global dipole field, giving preponderance to the higher-order multipole components \citep{metcalfe16}. As these components are not expected to be capable of efficiently draining angular momentum via winds \citep{see19}, in comparison to the global dipole geometry, it is argued that the rotational evolution of solar mass stars older than 4 Gyr would be driven only by structural variations (VS16). As seen in Fig. \ref{fig:rot_age}, HIP 102152 (red cross) fits properly to the scenario of smooth rotational evolution described in LO19 (see their Fig. 2), with no need to introduce any additional degree of freedom to emulate the angular momentum loss inefficiency around 4 Gyr (Ro$_{\rm crit}\sim$2). The shaded region in blue is the prediction band for smooth rotational evolution models of 0.97 $\pm$ 0.06 $M_\odot$ ($\pm2\sigma$). As \feh$\sim$ 0.00, we restricted our analysis to the solar metallicity tracks. LO19 used a sample of solar twins with measured \Prot and determined the most probable age for a magnetic transition, finding $\gtrsim8\, $Gyr or Ro$_{\rm crit}$ = $2.6^{+\infty}_{-0.1}$. These values are in line with M18 prediction. Therefore, stars of $\sim$1 solar mass travel along the main sequence without undergoing any dynamo disruption. For the sake of consistency, if this transition still occurs along the main sequence lifetime, a conservative lower limit within $2\sigma$ was estimated to be Ro$_{\rm crit}\gtrsim$ 2.3 (or $t_{\rm crit}\gtrsim$ 5.3 Gyr). We repeated the same procedure adopted in LO19 to derive the  probability distribution for Ro$_{\rm crit}$ yielding $\gtrsim$2.8 (or $t_{\rm crit}\gtrsim$9 Gyr). The only difference between our procedure and those from LO19 is that we are deriving the probability for a single star instead of a joint probability given by a sample of solar twins. 
\section{Conclusions}
We use high cadence HARPS observations to provide constraints to the possible solar magneto-rotational evolutionary path $\sim$ 3.5 Gyr from now when it is near to leave the main sequence. As the chromospheric modulations in old stars are very subtle, we approached this issue in multiple ways to detect the rotational signal. We derived several activity indices throughout the spectral coverage of HARPS and tested their sensitivities against the weak rotational signal of HIP 102152. To do so, quasi-periodic models based on Gaussian processes were used. We found that \ion{Ca}{II} K line stands out as the most sensitive indicator ($B_{\rm factor}\sim10^{70}$), followed by the \ion{Ca}{II} S index (H$+$K), and H$\epsilon$. For the best indicators, it was detected a \Prot of 36 $\pm$ 1.4 days with very high statistical significance. Both \tess\, photometry and projected rotational period (\Psini) are consistent with the spectroscopic indicators. These values are in full agreement with the expected rotation for an 8 Gyr-old, 1 solar mass, and solar metallicity star, in opposition to the weakened magnetic braking scenario from VS16. Thus, HIP 102152 supports the smooth rotational evolution of the Sun and validates the use of gyrochronology \citep{barnes07}.
%@arxiver{AGE_ROTATION.pdf}

\section*{Acknowledgements}
We thank the anonymous referee for useful comments that helped improve this manuscript. D.L.O is grateful to the tenacious Brazilian workers and taxpayers, whose effort enabled the existence of this project.
D.L.O. and J.M. thanks support from FAPESP (2016/20667-8; 2018/04055-8). G.P. acknowledges the support from CAPES, FAPESP, and MackPesquisa fundings. J.Y.G. acknowledges the support from CNPq.

\bibliographystyle{mnras}
%\bibliography{bibli2} % if your 
%%%%%%%%%%%%%%%%%%%%%%%%%%%%%%%%%%%%%%%%%%%%%%%%%%

% Don't change these lines
\bsp	% typesetting comment
\label{lastpage}
\end{document}